
\font\subtit=cmr12
\font\name=cmr8

\def\plb#1#2#3#4{#1, {\it Phys. Lett.} {\bf {#2}}B (#3), #4}
\def\npb#1#2#3#4{#1, {\it Nucl. Phys.} {\bf B{#2}} (#3), #4}

\def\cmp#1#2#3#4{#1, {\it Comm. Math. Phys.} {\bf {#2}} (#3), #4}
\def\lmp#1#2#3#4{#1, {\it Lett. Math. Phys.} {\bf {#2}} (#3), #4}

\def\mpl#1#2#3#4{#1, {\it Mod. Phys. Lett.} {\bf A{#2}} (#3), #4}
\def\ijmpa#1#2#3#4{#1, {\it Int. Jour. Mod. Phys.} {\bf A{#2}} (#3), #4}

\input harvmac
\def\UWrLMU#1#2#3#4
{\TITLE{U.T.F. #1} {IFT UWr #2/\number\yearltd}{#3}{#4}}
\def\TITLE#1#2#3#4{\nopagenumbers\abstractfont\hsize=\hstitle\rightline{#1}
\vskip 1pt\rightline{#2} \vskip 1in \centerline{\subtit #3} \vskip 1pt
\centerline{\subtit #4}\abstractfont\vskip .5in\pageno=0}
\UWrLMU{343}{886} {OPERATOR FORMALISM FOR b--c SYSTEMS WITH
$\lambda=1$}{ON GENERAL ALGEBRAIC CURVES}
\centerline{F.  F{\name ERRARI}$^{a}$, J.  S{\name OBCZYK}$^b$}
\smallskip $^a${\it Dipartimento di Fisica,
Universit\`a di Trento, 38050 Povo (TN), Italy and INFN, Gruppo
Collegato di Trento, E-mail:  francof@galileo.science.unitn.it
}
\smallskip $^b${\it Institute for Theoretical Physics, Wroc\l aw
University, pl.  Maxa Borna 9, 50205 Wroc\l aw, Poland, E-mail:
jsobczyk@proton.ift.uni.wroc.pl}\smallskip
\vskip 2cm
\centerline{ABSTRACT}
{\narrower
In this letter we develope an operator formalism for the
$b-c$ systems with conformal weight $\lambda=1$ defined on a general
closed and orientable Riemann
surface. The advantage of our approach is that the Riemann surface is
represented as an affine algebraic curve. In this way it is possible
to perform
explicit calculations in string theory at any perturbative order.
Besides the obvious applications in string theories and conformal
field theories, (the $b-c$ systems at $\lambda=1$
are intimately related to the free scalar field theory), the operator
formalism presented here sheds some light also on the quantization of
field theories on Riemann surfaces. In fact, we are able to
construct explicitly the vacuum state of the $b-c$ systems and to
define creation and annihilation operators. All the amplitudes are
rigorously computed using simple normal ordering prescriptions as in
the flat case.
}
\Date{January, 1995}
\newsec { INTRODUCTION}
\vskip 1cm
In a recent paper
\ref\feso{F. Ferrari and J. Sobczyk {\it
Operator Formalism on General Algebraic Curves}, preprint U.T.P. 333,
IFT UWr 879/94.}
we have introduced an operator formalism for the $b-c$
systems with integer conformal weight $\lambda\ge 2$
on general Riemann surfaces.
In our construction
the fields are expanded in generalized Laurent series. Upon quantization,
the coefficients of the series become either creation and annihilation
operators or correspond to zero modes of the theory.
The advantage of the approach used in \feso\ is
that the Riemann
surfaces are represented as algebraic curves, i.e. as
$n-$sheeted coverings of the complex
plane. This allows very explicit calculations and a better
understanding of free field theories defined on nontrivial two
dimensional topologies.
A remarkable result which we obtain is the proof that the Hilbert
space of physical states is splitted into a finite number of "independent"
spaces. With all these ingredients we have been able to calculate
in an explicit way the correlation functions of the $b-c$ systems.
\vskip 1pt
As might be expected, the main difficulty of the operator formalism is
to deal properly with the
zero modes. For that reason, in \feso\ we have restricted
ourselves
to the case $\lambda\geq 2$, where only the zero modes of the $b$ fields are
present. In this paper we are going to generalize our approach
also to the interesting case in which $\lambda=1$.
With respect to ref. \feso, the calculations at $\lambda=1$ are
complicated by the presence of the $c$ zero mode, which modifies the
definition of the vacuum state and of the normal ordering of the
fields.
\vskip 1pt
We notice that our formalism is valid on
very general algebraic curves
given by any Weierstrass polynomial that is nondegenerate. We also have
assumed that all the branch points lie in a finite region of the
complex plane.
These restrictions are adopted only to simplify the technical aspects of the
construction. In fact,
the crucial point of the reasoning
is the possibility of
expanding the Weierstrass kernel in terms of
multivalued modes
(elements of the generalized Laurent expansions) of the fields. This
possibility has been proven for arbitrary Weierstrass polynomials
in \feso. For instance, in
\feso\ we provide some examples showing that the way in which we introduce
the operator formalism is entirely general. Let us mention also that the
problem of constructing operator formalism for $b-c$ systems on Riemann
surfaces has been discussed by other authors but in a
different and less explicit (from the point of view of physical applications)
mathematical framework
\ref\raina{\cmp{A.  K.  Raina}{122}{1989}{625};
{\it ibid.}  {\bf 140} (1991), 373; {\it Lett.  Math.  Phys.}
{\bf 19} (1990), 1; {\it Expositiones Mathematicae} {\bf 8} (1990),
227; {\it Helvetica Physica Acta} {\bf 63} (1990), 694; A review of an
algebraic geometry approach to a model quantum field theory on a
curve, based on the invited talks in the conferences:  {\it Topology
of Moduli Space of Curves}, Kyoto, September 1993, {\it Vector Bundles
on Curves - New Directions}, Bombay and Madras, December 1993, {\it
International Colloquium on Modern Quantum Field Theory II} Bombay,
January 1994.}
\vskip 1pt
There are various applications of our results.
First of all, the new formalism should be useful in perturbative string
theory, generalizing to any Riemann surface previous computations
performed in the special case of hyperelliptic curves
\ref\hyper{\npb{E. Gava, R.  Iengo and C.  J.  Zhu}{323}{1989}{585};
\plb{E.  Gava, R.  Jengo and G.  Sotkov}{207}{1988}{283};
\plb{R.  Jengo and C.  J.  Zhu}{212}{1988}{313};
\npb{D.  Lebedev and A.  Morozov}{302}{1986}{163};
\plb{A.  Yu.  Morozov and A.  Perelomov}{197}{1987}{115};
\npb{D.  Montano}{297}{1988}{125}.}.
Moreover, there has been in recent times some interest in
the quantization of two dimensional free field theories
immersed in curved space-times \ref\curst{D. J. Lamb, A. Z. Capri and
S. M. Roy, {\it Massive particle creation in a static $1+1$
dimensional spacetime}, Preprint Alberta Thy-1-93, hep-th/9411225; D.
J. Lamb and A. Z. Capri, {\it Finite particle creation in $1+1$ dim.
compact in space}, Preprint Alberta Thy-25-94, hep-th/9412010; \cmp{A. L.
Carey, M. G. Eastwood and K. C. Hannabus}{130}{1990}{217}; N. B.
Birrell and P. C. W. Davies, {\it Quantum Fields in Curved Space},
Cambridge University Press, (1982).}.
 From this point of view, the example of the $b-c$ systems,
in which the vacuum state is
explicitly constructed and the amplitudes can be computed using simple
normal ordering prescriptions, is remarkable. As a matter of fact,
the Riemann
surfaces are not globally hyperbolic manifolds and, as a consequence,
there is not an
unique definition of time \ref\fulling{S. A. Fulling, {\it Aspects of
Quantum Field Theory in Curved Space-Time}, Cambridge University
Press, New York 1989.}. Nevertheless, we are able to show that it is
still possible to define creation and annihilation operators,
corresponding to the creation and destruction of certain
multivalued modes.
Other applications are related to the representation of the solutions
of deformed Kniznik-Zamolodchikov equations
\ref\cmph{\cmp{F.  Ferrari}{156}{1993}{179}.}\nref\pp{\mpl{S.
Pakuliak and A.
Perelomov}{9}{1994}{1791}.}--\ref\fas{
\cmp{F.  A.  Smirnov}{155}{1993}{459}.},
integrable models
\ref\matve{V.  Matveev, {\it Deformations of
Algebraic Curves and Integrable Nonlinear Evolution Equations},
Preprint MPI-PTH/14-91.}--\ref\matvee{D.  A.  Korotkin and V.  B.
Matveev, {\it Leningrad Math.  Jour.}  {\bf 1} (1990), 379.}
and relation between CFT on the complex plane and $b-c$ systems on Riemann
surfaces
\ref\knirev{\cmp{V.  G.  Knizhnik}{112}{1987}{587};
{\it Sov.  Phys.  Usp.}  {\bf 32}(11) (1989) 945.}\nref\brzn{\ijmpa{M.
A.  Bershadsky and A.  O.  Radul}{2}{1987}{165}.}\nref\plbff{\plb{F.
Ferrari}{277}{1992}{423}.}\nref\ijmp{\ijmpa{F. Ferrari}{9}{1994}
{313}.}--\ref\fsu{F.  Ferrari,
J.  Sobczyk and W.  Urbanik, {\it Operator
Formalism on the $Z_n$ Symmetric Algebraic Curves}, Preprint LMU-TPW
93-20, ITP UWr 856/93.}.
\vskip 1cm
\newsec{GENERALIZED LAURENT EXPANSIONS ON ALGEBRAIC CURVES}
\vskip 1cm
We consider a theory of free fermionic $b - c$ fields with spin
$\lambda=1$ defined by the following action
\eqn\action{S_{\rm
bc}=\int_{\Sigma_g}d^2\xi\left(b\bar\partial c+{\rm c.c.}\right).}
$\xi$ and $\bar \xi$ are complex coordinates on a Riemann surface $\Sigma_g$
described by means of an algebraic equation in {\bf CP}$_2$
\ref\cenr{F.  Enriques and O.  Chisini, {\it Lezioni sulla Teoria
Geometrica delle Equazioni e delle Funzioni Algebriche}, Zanichelli,
Bologna (in italian).}--\ref\grha{P. Griffiths
and J. Harris, {\it Principles of Algebraic
Geometry}, John Wiley \& Sons, New York 1978.}:
\eqn\curve{F(z,y) = P_n(z)y^n + P_{n-1}(z)y^{n-1} +
\ldots + P_1(z)y + P_0 = 0.}
The $P_s(z)=\sum\limits_{m=0}^{n-s}\alpha_{s,m}z^m$ are
polynomials in $z$ of degree at most $n-s$, $s=0,\ldots,n$ and the
$\alpha_{s,m}$ are complex parameters.  $y$ can be viewed either as a
multivalued function on the complex sphere {\bf CP}$_1$ or as a
meromorphic function on the Riemann surface defined by eq.  \curve. In both
cases we can think of $\Sigma_g$ as of an $n$-branched
covering of {\bf CP}$_1$,
with the function $y$ taking (in general)
$n$ different values on the $n$ branches forming the Riemann surface.
The function $y(z)$ has poles only at $z=\infty$.
In order to fix the ideas (see the discussion in \feso ),
we require the algebraic curve \curve\
to be nondegenerate \grha . Then
the genus of $\Sigma_g$ is $g={(n-1)(n-2)\over 2}$. All the other Riemann
surfaces can be obtained by some limiting procedure in the parameters
$\alpha_{s,m}$. In \feso\ we argue that our operator formalism is kept
untouched while performing such limits.
\vskip 1pt
The solutions of
the classical equations of motion descending from eq.  \action\
can be represented as
\eqn\bdz{b(z)=\sum\limits_{k=0}^{n-1}\sum\limits_{i=-\infty}
^\infty b_{k,i}z^{-i-1}f_k(z)}
\eqn\cdz{c(z)=
\sum\limits_{k=0}^{n-1}\sum\limits_{i=-\infty}^\infty
c_{k,i}z^{-i}\phi_k(z)}
with $f_k$ and $\phi_l$ chosen as follows ($k,l=0, ..., n-1$):
\eqn\fkn{ f_k(z) = {y^{n-1-k}(z)dz\over F_y(z,y(z)) } }
\eqn\phikn{\phi_l(w)  =
y^l(w)+y^{l-1}(w)P_{n-1}(w)+y^{l-2}(w)P_{n-2}(w)+...+P_{n-l}(w) }
We notice that we have introduced two different expansions for the
fields $b$ and $c$. This is only for the sake of convenience in the
formulation of the operator formalism.\vskip 1pt
In order to check the absence of spurious singularities in the
amplitudes, we shall also need the following divisors:
\eqn\divdz{ [dz]=\sum^{n_{bp}}_{p=1}(\nu_p-1) a_p\ -\
2\sum^{n-1}_{j=0}\infty_j } \eqn\divy{ [y]=\sum^n_{r=1}q_r\ -\
\sum^{n-1}_{j=0}\infty_j } \eqn\divdf{
[F_y]=\sum^{n_{bp}}_{p=1}(\nu_{p}-1)a_p\ -\
(n-1)\sum^{n-1}_{j=0}\infty_j }
In eq.  \divy\ the
$a_p$ are the branch points of the curve $\Sigma_g$ determined by the
condition \grha:
\eqn\brpo{F(z,y) = F_y(z,y) = 0}
We suppose that there are $n_{bp}$ finite branch points
of multiplicity $\nu_s$, where $\nu_s$ describes the number of
branches of $y$ connected at the branch point $a_s$.
The $q_j$ denote the
zeros of $y$ which, when projected from the Riemann surface on the $z$
complex plane, coincide with the zeros of $P_0(z)$.  Moreover,
$\infty_j$ describes the projection of the point at infinity on the
$j$-th sheet of the Riemann surface.  Finally, in our conventions
positive and negative integers denote the order of the zeros and of
the poles respectively.  Starting from eqs.  \divdz-\divdf, it is also
possible to find the divisors of the differentials \fkn:
\eqn\divf{ [z^if_k]=(n-1-k)\sum^n_{s=1}q_s + i\sum^{n-1}_{l=0} 0_l
+(k-2-i)\sum^{n-1}_{l=0}\infty_l }
where, using
the same notation exploited for the point at infinity, $0_l$ denotes the
projection of the point $z=0$ on the $j-th$ sheet.
\vskip 1pt
Next we explicitly derive the form of
the zero modes (holomorphic differentials)
associated to the equations of motion.  We try
the following ansatz:
\eqn\zm{\Omega_{k,i} = f_k(z)z^{-i-1} }
 From \divf\ it turns out that
$\Omega_{k,i}$ has no singularities whenever
\eqn\zma{ i\leq -1\quad\quad {\rm and}\quad\quad k-1+i \geq 0 }
Skipping the simple cases
corresponding to genus zero and one Riemann surfaces ($n=2,3$), we
obtain the conditions determining the $g$
independent zero modes of the form \zm\ :
\eqn\casethree{\cases{k=2,3,\ldots,n-1\cr -k+1\le i\le -1\cr}} while
\eqn\nbkthree{N_{b_k}=k-1\qquad\qquad\qquad k=2,\ldots,n-1}
\vskip 1pt
Obviously, the overall number of $b$ zero modes is equal to
${(n-1)(n-2)\over 2}=g$.
Looking at \phikn\ we also find out that the
$c$ zero mode corresponds to $l=0$.
\vskip 1cm
\newsec{THE OPERATOR FORMALISM}
\vskip 1cm
In the previous section we have shown how to identify the classical degrees
of freedom as coefficients $b_{i,k}$ and $c_{i,k}$.
Now we perform the quantization transforming $b_{i,k}$ and
$c_{i,k}$ into creation and annihilation operators.
We divide the degrees of freedom of
the $b-c$ systems into $n$ sectors, numbered by the index
$k=0,\ldots,n-1$ and characterized by the tensors $f_k$ for the $b$
fields and by the tensors $\phi_k$ for the $c$ fields.
We assume that in our operator
formalism the modes $z^jf_k(z)$ labelled by different indices $k$ do
not interact.  This hypothesis, to be proven a posteriori, implies
that the space on which the $b-c$ fields defined on a Riemann surface
act can be decomposed into a set of $n$ independent Hilbert spaces if
the Riemann surface is represented as an $n-$sheeted branch covering
over {\bf CP}$_1$.\smallskip
In order to set up the operator formalism, we
postulate the following commutation relations:
\eqn\commrel{\{b_{k,i},c_{k',i'}\}=\delta_{kk'}\delta_{i+i',0}.}
Moreover, we define $n$ vacua $|0\rangle_k$, $k=0,\ldots,n-1$, in
such a way that the modes with negative powers of $z$ are annihilation
operators
\eqn\ban{b^-_{k,i}|0\rangle_k\equiv
b_{k,i}|0\rangle_k=0
\qquad\qquad\qquad\left\{\eqalign{k=&0,\ldots,n-1\cr i \ge&
0\cr}\right.}
\eqn\can{c^-_{k,i}|0\rangle_k\equiv
c_{k,i}|0\rangle_k=0
\qquad\qquad\qquad\left\{\eqalign{k=&0,\ldots,n-1\cr i\ge&
1\cr}\right.}
Naively, the corresponding right vacua are given by:
\eqn\bcreaa{{}_k\langle0| b^+_{k,i}\equiv {}_k\langle 0|b_{k,i}=0
\qquad\qquad\qquad\left\{\eqalign{k=&2,\ldots,n-1\cr i \le&
-k\cr}\right.}
\eqn\bcreab{{}_k\langle0| b^+_{k,i}\equiv
{}_k\langle 0|b_{k,i}=0 \qquad\qquad\qquad\left\{\eqalign{k=&0,1\cr i
\le& -1\cr}\right.}
\eqn\ccreaa{{}_k\langle 0|c^+_{k,i}\equiv
{}_k\langle 0|c_{k,i}=0
\qquad\qquad\qquad\left\{\eqalign{k=&1,\ldots,n-1\cr i\le&
0\cr}\right.}
\eqn\ccreab{{}_0\langle 0|c^+_{0,i}\equiv {}_0\langle
0|c_{0,i}=0 \qquad\qquad\qquad\left\{\eqalign{k=&0\cr i\le&
-1\cr}\right.}
where $N_{b_k}=k-1$ from eq.  \nbkthree.  In order to
treat properly the zero modes, however, we have to introduce modified
vacua $_k\langle0'|$ defined by:
\eqn\modd{_k\langle0'|=_k\langle0|b_{k,-k+1}\ldots b_{k,-1}\qquad\qquad\qquad
k=2,\ldots,n-1}
\eqn\modvac{_1\langle0'|=_1\langle0|}
\eqn\moddd{_0\langle0'|=_0\langle0|c_{0,0}}
This is the generalization on Riemann surfaces of the recipe
given in the flat case in order to treat the zero modes (see ref.
\ref\fms{\npb{D. Friedan, E. Martinec and S.
Shenker}{271}{1986}{93}.}).
The modified vacua are normalized as follows:
\eqn\vaccond{_k\langle0'|0\rangle_k=1\qquad\qquad\qquad
k=0,\ldots,n-1}
Only the amplitudes containing at least a number
$N_{b_k}$ of $b$ fields in the sectors $k=2,\ldots,n-1$ and one $c$
field in the sector $k=0$ do not vanish.  The total vacuum $|0\rangle$
is given by:
\eqn\totvacket{|0\rangle=\prod_{k=0}^{n-1}|0\rangle_k}
\eqn\totvacbra{\langle0|=\prod_{k=0}^{n-1}{}_k\langle0|}
The following notations will be useful in the future:
\eqn\bkdz{b_k(z)=f_k(z) \sum\limits_{i=-\infty}^{\infty}b_{k,i}
z^{-i-1}} \eqn\ckdz{c_k(z)=\phi_k(z) \sum\limits_{i=-\infty}^\infty
c_{k,i}z^{-i}}
\eqn\bdzcdz{b(z)=\sum_{k=0}^{n-1}b_k(z)\qquad\qquad\qquad
c(z)=\sum_{k=0}^{n-1}c_k(z)}
Eq.  \bdzcdz\ is equivalent to the
decompositions \bdz\ and \cdz.  The only difference is that now the
$b_{k,i}$ and $c_{k,i}$ are operators.
\vskip 1pt
After the identification of $b_{i,k}$ and $c_{i,k}$ as creation
or annihilation operators (see \ban -- \ccreab ) it is possible
to introduce a natural
notion of normal ordering. The only modification comes from the fact that
some
$b_{i,k}$ as well as $c_{0,0}$ operators are neither annihilation nor
creation ones. According to eqs. \modd-\moddd,
we define the normal ordering by requiring that the operators
corresponding to
zero modes stand to the left of the annihilation operators.
Exploiting the commutation relations \commrel\
and the conventions established above, it is easy to see that the
relation between normal ordered and not ordered products of $b-c$
fields is given by:
\eqn\normord{b_k(z)c_k(w)=:b_k(z)c_k(w):+{1\over z-w}f_k(z)\phi_k(w)}
The ``time ordering'' is implemented by the
requirement that the fields $b(z)$ and $c(w)$ are radially ordered
with respect to the variables $z$ and $w$.
\vskip 1pt
We are now ready to compute the correlation functions of the $b-c$ systems.
Using eq.  \vaccond\ the proof of the following identity is
straightforward
\fsu, \feso:
\eqn\propone{{}_k\langle 0|b_k(z_1)\ldots
b_k(z_{N_{b_k}})|0\rangle_k= {\rm Det}\left|
\Omega_{k,j}(z_i)\right|\qquad\qquad\qquad \cases{k=2,\ldots,n-1\cr
i,j=1\ldots,N_{b_k}\cr}}
The zero modes $\Omega_{k,j}(z)$ can be expressed
in terms of $z$ and $y$ from eqs.  \fkn\ and \zm:
\eqn\zmexp{\Omega_{k,j}(z)={y^{n-k-1}z^{j-1}\over
F_y(z,y)}dz\qquad\qquad\qquad\cases{k=2,\ldots,n-1\cr
j=1,\ldots,k-1\cr}}
With the help of \propone\ it is possible to compute the correlator:
\eqn\sfunct{S(u;z_1,\ldots,z_{N_b})=\langle0|c(u)b(z_1)\ldots
b(z_{N_b})|0\rangle={\rm det}|\Omega_I(z_J)| }
Since $S(u;z_1,\ldots,z_{N_b})$ does not depend on the
coordinate $u$ of the $c$ zero mode, the following notation is
more convenient: $S(u;z_1,\ldots,z_{N_b})\equiv S(z_1,\ldots,z_{N_b})$.
In \sfunct\ $I,J=1,\ldots,N_b$ and $N_b$ denotes the total
number of zero modes. The $\Omega_J(z)$ represent all the
possible zero modes in the $b$ fields:
$$\Omega_I(z)\in\left\{\Omega_{k,i}(z)|1\le i\le N_{b_k},\qquad 2\le
k\le n-1\right\}$$
To demonstrate \sfunct, we rewrite the correlator
\sfunct\ as follows:
\eqn\expanded{S(z_1,\ldots,z_{N_b})=\sum_{s=0}^{n-1}
\sum_{r_0,\ldots,r_{n-1}\atop r_0+\ldots+r_{n-1}=N_b}\sum_\sigma{\rm
sign}
(\sigma)\prod_{k=0}^{n-1}{}_k\langle0|\prod_{l_k=\alpha(k)}^{\beta(k)}
b_k(z_{\sigma(l_k)})c_s(w)|0\rangle_k} where:
$$\alpha(0)=1\qquad\qquad\qquad\beta(0)=r_0$$
$$\cases{\alpha(k)=1+
\sum\limits_{m=0}^{k-1}r_m\cr \beta(k)=\alpha(k)+r_k-1\cr}\qquad
\qquad\qquad k=1,\ldots,n-1$$
$$\sigma(\alpha(k))\le\ldots\le\sigma(\beta(k))\qquad\qquad\qquad
k=0,\ldots, n-1$$
In eq.  \expanded\ the products of the fields $b$
and $c$ have been expanded in their components $b_k(z)$ and $c_k(z)$,
$k=0,\ldots,n-1$.  It is easy
to see that any contraction of a $b$ field with $c_s(w)$ leads to a
vanishing amplitude due to \vaccond.  For that reason, the only
nonvanishing amplitudes are those for which:  $$r_0=r_1=0,\qquad\qquad
r_k=N_{b_k}$$ and $s=0$.  Exploiting also eq.  \propone, we arrive at
the final result:
\eqn\finform{S(z_1,\ldots,z_{N_b})=\sum_\sigma{\rm sign}(\sigma)
\prod_{k=2}^{n-1}{\rm
det}\left[\Omega_{k,i_k}(z_{\sigma(l_k)})\right]} where
$i_k=1,\ldots,N_{b_k}$ and $l_k=1+\sum\limits_{m=2}^{k-1}N_{b_k},
\ldots,\sum\limits_{m=2}^kN_{b_k}$.  The right hand side of \finform\
is exactly the determinant
${\rm det}\left|\Omega_I(z_J)\right|$. This way of expanding the
determinants is explained in \ref\detnl{F. Neiss and H. Liermann, {\it
Determinanten und Matrizen}, $8^{\rm th}$ Ediction, Springer Verlag 1978,
in german.}.
Thus eq.  \sfunct\ has been
demonstrated. More details of the calculation
are contained in ref. \feso.
\smallskip
The next amplitude to be computed is the propagator $G(z,w)dz$.
\eqn\propdef{G(z,w)dz\equiv{\langle0|b(z)c(w)c(u)
\prod\limits_{I=1}^{N_b}b(z_I)|0\rangle\over
\langle0|c(u)
\prod\limits_{I=1}^{N_b}b(z_I)|0\rangle
}}
 From eq. \normord\ the normal ordering between any two fields $b$ and
$c$ becomes:
\eqn\totnormord{b(z)c(w) = :b(z)c(w): + K(z,w)dz}
where
\eqn\kl{K(z,w)dz=\sum_{k=0}^{n-1}{f_k(z)\phi_k(w)\over z-w}}
Using the Wick theorem we find:
$$G(z,w)dz=K(z,w)dz-K(z,u)dz+$$
$$\sum\limits_{J=1}^{N_b}(-1)^JK(z_J,w)dz_J{S(z_1,\ldots,
z_{J-1},z,z_{J+1}\ldots,z_{N_b})\over S(z_1,\ldots,z_{N_b})} -$$
\eqn\pddf{\sum\limits_{J=1}^{N_b}(-1)^JK(z_J,u)dz_J{S(z_1,\ldots,
z_{J-1},z,z_{J+1},\ldots,z_{N_b})\over S(z_1,\ldots,z_{N_b})}}
The correlators $S(z_1,\ldots, z_{J-1},z,z_{J+1},\ldots,z_{N_b})$
can be evaluated by means of eq. \sfunct .
Let us introduce now the third kind differential \ref\tkd{\ijmpa{F.
Ferrari} {5}{1990}{2799}.}
$$\omega_{wu}(z)dz=K(z,w)dz-K(z,u)dz =$$
\eqn\tkdiff{{F(w,y^{(l)}(z))\over (y^{(l)}(z)-y^{(r)}(w))F_y(z,y^{(l)}(z))}
{dz\over z-w}-
{F(u,y^{(l)}(z))\over (y^{(l)}(z)-y^{(s)}(u))F_y(z,y^{(l)}(z))}
{dz\over z-w}}
$\omega_{wu}(z)dz$ is a differential in $z$ with simple poles at points
$z=w$ and $z=u$ (it should be understood that by point $u$ or $w$ we mean
a point on a definite sheet of $y$, namely $r$ for $w$ and $s$ for $u$, while
the point $z$ lies on the sheet $l$). With these conventions
eq. \pddf\ becomes:
\eqn\pluo{G(z,w)dz={{\rm det}
\left|\matrix{\omega_{wu}(z)&\Omega_1(z)&\ldots&\Omega_{N_b}(z)\cr
\omega_{wu}(z_1)&\Omega_1(z_1)&\ldots&\Omega_{N_b}(z_1)\cr
\vdots&\vdots&\ddots&\vdots\cr
\omega_{wu}(z_{N_b})&\Omega_1(z_{N_b})&\ldots
&\Omega_{N_b}(z_{N_b})\cr}\right|
\over |{\rm det}\Omega_I(z_J)|}}
It can be verified that the propagator \pluo\ does not contain spurious poles.
Further application of the Wick theorem enables us to calculate more
complicated correlation functions.
$$G_{NM}(z_1\ldots z_N;w_1\ldots w_M)=$$
\eqn\gnm{
{{\rm det}
\left|\matrix{\omega_{w_1w_M}(z_1)&\ldots&\omega_{w_{M-1}w_M}(z_1)&
\Omega_1(z_1)&\ldots&\Omega_{N_b}(z_1)\cr
\omega_{w_1w_M}(z_2)&\ldots&\omega_{w_{M-1}w_M}(z_2)&
\Omega_1(z_2)&\ldots&\Omega_{N_b}(z_2)\cr
\vdots&\ddots&\vdots&\vdots&\ddots&\vdots\cr
\omega_{w_1w_M}(z_N)&\ldots&\omega_{w_{M-1}w_M}(z_N)&
\Omega_1(z_N)&\ldots&\Omega_{N_b}(z_N)\cr}\right|
\over |{\rm det}\Omega_I(z_J)|}}
The way in which \gnm\ is written suggests that the point $w_M$ is
treated in a special way. It is however only apparent due to the formula
\tkdiff . In fact \gnm\ has all the necessary properties of a "good"
correlation function
\ref\vv{\npb{E.  Verlinde and H.  Verlinde}{288}{1987}{357};
\ijmpa{M.  Bonini and R.  Iengo}{3}{1988}{841}.}.
\vskip 1cm
\newsec{FINAL REMARKS}
\vskip 1cm
We have shown that the new operatorial formalism on Riemann surfaces
proposed in \feso\ can be applied also to more complicated cases in which
both the $b$ and $c$ zero modes are present. All information
about the zero modes
is absorbed in the suitable definition of the modified vacuum state of
the
theory. With the prescriptions given here it is possible to compute
the amplitudes of the $b-c$ systems at $\lambda=1$.
Due to the lack of space some details concerning the Wick
theorem and the application of the operator formalism to the scalar
fields have been omitted.
These topics will be treated in a forthcoming paper, where
we will also derive
in an explicit way all the formulas required in the perturbative
formulation of the string theory \ref\fsinprep{F. Ferrari and J.
Sobczyk, {\it Explicit Formulas of String theory}, in preparation.}.
\listrefs
\end

\ref\fms{\npb{D.  Friedan, E.  Martinec and S.  Shenker}{271}{1986}{93}}
\ref\ceh{\cmp{A.  L.  Carey, M.  G.  Eastwood and
K.  C.  Hannabus}{130}{1990}{217}}
\ref\ffstr{\ijmpa{F.  Ferrari}{A5}{1990}{2799}}
\ref\janstr{\lmp{J.  Sobczyk and W.  Urbanik}{21}{1991}{1};
\mpl{J.  Sobczyk}{A6}{1991}{1103}; {\it ibid.}  {\bf A8} (1993), 1153.}
\ref\fay{J.  D.  Fay, Theta Functions on Riemann Surfaces, Lecture
Notes in Mathematical Physics no.  352, Springer Verlag, 1973.}
\ref\eo{\plb{T.  Eguchi and H.  Ooguri}{187}{1987}{127}}
\ref\abmn{\cmp{L.  Alvarez-Gaum\'e, J.-B.
Bost, G.  Moore, P.  Nelson and C.  Vafa}{112}{1987}{503}}
\ref\agr{L.  Alvarez-Gaum\'e, C.  Gomez and
C.  Reina, New Methods in String Theory, in:  Superstrings '87, L.
Alvarez-Gaum\'e (ed.), Singapore, World Scientific 1988; N.  Kawamoto,
\cmp{Y.  Namikawa, A.  Tsuchiya and Y.  Yamada}{116}{1988}{247}}
\ref\blmr\cmp{L.  Bonora, A.  Lugo, M.  Matone
and J.  Russo}{123}{1989}{329}}
\ref\semi{\plb{A.  M.  Semikhatov}{212}{1988}{357};
\plb{P.  di Vecchia}{248}{1990}{329};
\plb{O.  Lechtenfeld}{232}{1989}{193};
\plb{U.  Carow-Watamura, Z.  F.  Ezawa, K.  Harada, A.  Tezuka and S.
Watamura}{227}{1989}{73}}